\documentclass{article}[12pt]
\usepackage{setspace}
\usepackage{graphicx}
\usepackage{xcolor}

\title{Testing for an ignorable sampling bias under random double truncation}
\author{Jacobo de U\~na-\'Alvarez\\jacobo@uvigo.es \\Department of Statistics and OR \& CINBIO,  Universidade de Vigo, Vigo, Spain}

\begin{document}

\maketitle

\begin{abstract}
In clinical and epidemiological research doubly truncated data often appear. This is the case, for instance, when the data registry is formed by interval sampling. Double truncation generally induces a sampling bias on the target variable, so proper corrections of ordinary estimation and inference procedures must be used. Unfortunately, the nonparametric maximum likelihood estimator of a doubly truncated distribution has several drawbacks, like potential non-existence and non-uniqueness issues, or large estimation variance. Interestingly, no correction for double truncation is needed when the sampling bias is ignorable, which may occur with interval sampling and other sampling designs. In such a case the ordinary empirical distribution function is a consistent and fully efficient estimator that \textcolor{black}{generally} brings remarkable variance improvements \textcolor{black}{compared to the nonparametric maximum likelihood estimator}. Thus, identification of such situations is critical for the simple and efficient estimation of the target distribution. In this paper we introduce for the first time formal testing procedures for the null hypothesis of ignorable sampling bias with doubly truncated data. The asymptotic properties of the proposed test statistic are investigated. A bootstrap algorithm to approximate the null distribution of the test in practice is introduced. The finite sample performance of the method is studied in simulated scenarios.  Finally, applications to data on onset for childhood cancer and Parkinson's disease are given. Variance improvements in estimation are discussed and illustrated.
\end{abstract}

\section{Introduction}
\label{sec:intro}

The issue of random double truncation is ubiquitous in clinical and epidemiological research, among other fields. Double truncation appears when the observation of the target variable is restricted by two random limits. An example is found in Survival Analysis, when the target is an event time, and the data correspond to all the events between two specific dates (Moreira and de U\~na-\'Alvarez, 2010). Such sampling design has been often referred to as interval sampling (Zhu and Wang, 2012). Double truncation can be regarded as an extension of left-truncation, a well-known issue that affects cross-sectional data or registries with delayed entries. Autopsy-confirmed diseases is a remarkable example of double truncation in such left-truncated settings, since undergoing the event of interest (death) before the end of the study becomes then a prerequisite for recruitment; this induces truncation from the right and, thus, results in doubly truncated event times (Rennert and Xie, 2019).

Unlike for one-sided (left or right) truncation, the nonparametric maximum likelihood estimator (NPMLE) under random double truncation \textcolor{black}{does not have a closed form}. Efron and Petrosian (1999) introduced the NPMLE of a cumulative distribution function \textcolor{black}{(CDF)} under random double truncation; the motivation was found in the analysis of quasar luminosities that are subject to random detection limits. These authors exploited the distinctive features of double truncation to introduce a self-consistency equation for Turnbull (1976)'s estimator and, consequently, to propose an iterative algorithm for its calculation. The Efron-Petrosian NPMLE was further investigated by Shen (2010), Moreira and de U\~na-\'Alvarez (2010), Emura et al. (2015) and, more recently, by de U\~na-\' Alvarez and Van Keilegom (2021), who developed \textcolor{black}{asymptotic theory} to cope with double truncation in \textcolor{black}{the presence of covariates}. On the other hand, several \texttt{R} packages implementing the NPMLE for doubly truncated data have been launched along the last decade, so the application of the Efron-Petrosian estimator has become easy; available packages include \texttt{DTDA} (Moreira et al., 2022), \texttt{SurvTrunc} (Rennert, 2018) and \texttt{double.truncation} (Emura et al., 2020). \textcolor{black}{Importantly, the consistency of the Efron-Petrosian estimator depends on the assumption of quasi-independence between the target variable and the truncating variables. Testing procedures for a quasi-independence assumption in this setting have been investigated; see for instance Martin and Betensky (2005). On the other hand, Moreira et al. (2021) introduced a copula-based extension of the Efron-Petrosian estimator to cope with possibly dependent truncation. Throughout this paper it is assumed that the truncating variables are independent of the target.}

The sampling bias induced by double truncation has been discussed and illustrated in a number of research papers. For instance, Zhu and Wang (2012) \textcolor{black}{investigated the sampling bias} when estimating the birth rate from cancer registries obtained with interval sampling. In particular, they found that the plausible linear trend for the birth process was converted into a rather unrealistic bell-shaped curve due to the truncation effects. Mandel et al. (2018) discussed the biases that may arise in the proportional hazards regression model when fitted from interval sampling data too. Specifically, when looking for a relationship between genetic information and age at onset for Parkinson's disease, they found substantial differences between the estimated coefficients with and without the correction for double truncation. The sampling bias for the Parkinson's disease study was further explored in de U\~na-\' Alvarez (2020a), who \textcolor{black}{concluded} the oversampling of patients with intermediate ages at diagnosis. Rennert and Xie (2019) described the issue in the setting of autopsy-confirmed neurodegenerative diseases. Interestingly, they illustrated how by ignoring the double truncation one may overestimate or underestimate the target survival, depending on the particular situation. The impact of the sampling bias in the Mann-Whitney two-sample test was pointed out by these authors too. On the contrary, Moreira and de U\~na-\'Alvarez (2010) discussed how, with interval sampling, the sampling bias may vanish when the truncating variables are uniformly distributed; this was indeed the case for the childhood cancer registry investigated in the referred paper.

To sum up, random double truncation may, or may not, induce a sampling bias on the target variable. It is convenient to identify in practice whether such a potential sampling bias exist, at least due to the following reasons:

\begin{itemize}
	\item When there is no sampling bias the empirical cumulative distribution function (ECDF) is a consistent (and efficient) estimator of the target CDF, so there is no need to look for alternative estimators;
	\item The NPMLE which corrects for double truncation may not exist, or may be non-unique (Xiao and Hudgens, 2019); and
	\item The \textcolor{black}{estimation} of a distribution from doubly truncated data given by the corresponding NPMLE, when it exists and is unique, usually entails a large variance.
\end{itemize}

\noindent In other words, when there is no sampling bias one will have a strong motivation to apply the ECDF instead of the Efron-Petrosian NPMLE. And this is why testing for an ignorable sampling bias under double truncation becomes relevant.

The potential sampling bias under double truncation can be assessed from the joint graphical display of the ECDF and the Efron-Petrosian NPMLE. When both curves are close together one gets support for the hypothesis of ignorable sampling bias. Similarly, one may plot the NPMLE for the sampling probability to check whether it is roughly constant on the support of the target variable; see for instance Example 2.1.4 in de U\~na-\' Alvarez et al. (2021). Formal testing methods that guarantee a given significance level are however missing in the literature at the time of writing. A related reference is Moreira et al. (2014), who introduced and investigated through simulations several statistics to test for a parametric model for the \textcolor{black}{pair of truncating variables}. Still, a key difference in the current setting is that the null hypothesis of ignorable sampling bias does not characterize the (bivariate) truncation distribution; see Section 2 for further details. This complicates the application of the ideas in Moreira et al. (2014) to test for no sampling bias. In particular, a new resampling plan to approximate the null distribution of the introduced test statistic will be needed.

The interest in the random double truncation model has rapidly increased in the last years. Recent methods to handle doubly truncated outcomes include, among other topics, smoothing methods (Moreira and Van Keilegom, 2013; Moreira et al., 2016), proportional hazards regression (Mandel et al., 2018; Rennert and Xie, 2018), rank regression for linear models (Ying et al., 2020), competing risks (de U\~na-\' Alvarez, 2020b), two-sample problems (Shen, 2013), the estimation of a bivariate distribution (Zhu and Wang 2012, 2014, 2015), or maximum likelihood theory for parametric models (Emura et al., 2017). \textcolor{black}{Formally testing for an ignorable sampling bias is important in all these settings with double truncation} \textcolor{black}{because, when there is no sampling bias, ordinary methods apply and the estimation variance can be reduced.}


The rest of the paper is organized as follows. In Section 2 we introduce the required notation and a test statistic for the null hypothesis of ignorable sampling bias. The proposed test is defined as the supremum distance between the Efron-Petrosian NPMLE and the ECDF of the doubly truncated outcomes. The asymptotic null distribution of the test statistic is obtained, and a bootstrap algorithm to approximate the null distribution in practice is proposed. In Section 3 a simulation study to investigate the finite sample performance of the proposed test is conducted. In Section 4 illustrative applications of the proposed methods to data on childhood cancer and Parkinson's disease are given. A final discussion that mentions some possible extensions of the introduced methods and alternative testing approaches for an ignorable sampling bias is given in Section 5.

\section{Methods}
\label{sec2}

\subsection{Notation and preliminary remarks}

Let $X$ be the target variable with CDF $F$, assumed to be supported on an interval $[a_X,b_X] \subset \mathcal{R}$ or, more generally, on a set $\mathcal{S}_X$ with lower and upper limits $a_X$ and $b_X$. Let $(U,V)$ be \textcolor{black}{a couple of truncating variables independent of $X$} with (bivariate) CDF $K$, so the triplet $(X,U,V)$ is observed only when $U\leq X\leq V$. It is assumed that $\alpha \equiv P(U\leq X\leq V)$ is strictly positive. In this setting, the \textcolor{black}{observation of $X$ is limited by the condition} $a_U\leq X\leq b_V$, where $a_U$ and $b_V$ are the lower and upper endpoints of the supports of $U$ and $V$, respectively. Then, it is natural to impose the restrictions $a_U\leq a_X$ and $b_X\leq b_V$ (Woodroofe, 1985). \textcolor{black}{If these conditions are violated, no consistent estimation of $F$ can be provided.} Similarly, one must assume $P(U\leq V)=P(U\leq b_X)=P(V\geq a_X)=1$ for the identifiability of $K$.

The sampling information is a set of independent and identically distributed (iid) observations $(X_i,U_i,V_i)$, $1\leq i\leq n$, with the conditional distribution of $(X,U,V)$ given $U\leq X\leq V$. In general, the ECDF of the $X_i$'s, $F_n^*(x)=n^{-1}\sum_{i=1}^n I(X_i\leq x)$, is not consistent for $F(x)$. This is because the CDF of $X_1$ is given by

\begin{equation}
	F^*(x)=\alpha^{-1}\int_{a_X}^x G(t)dF(t)
	\label{Fstar}
\end{equation}

\noindent with $G(t)=P(U\leq t\leq V)=\int_{u\leq t\leq v}dK(u,v)$ and, thus, it does not coincide with $F$ unless $G$ is constant. Note that the function $G$ reports the sampling probability for the values of the target $X$\textcolor{black}{; that is, $G(t)$ is the probability of observing $X=t$}. When $G$ is strictly positive on $\mathcal{S}_X$, (\ref{Fstar}) admits the reverse formulation

\begin{equation}
	F(x)=\alpha \int_{a_X}^x G(t)^{-1}dF^*(t)
	\label{Freverse}
\end{equation}

\noindent and $\alpha=1/\int_{a_X}^{b_X}G(t)^{-1}dF^*(t)$ holds. \textcolor{black}{Assumption (C1) below ensures that $G$ is \textcolor{black}{strictly positive, provided that the densities of the truncating variables exist and have convex support}; this implies in particular that there are no regions within $[a_U,b_V]$ where $F$ is not identifiable.} Equation (\ref{Freverse}) reveals that $F$ can be estimated from $F_n^*$ and a consistent estimator for $G$. Actually, this is the classical idea behind inverse-probability-weighting estimation, where each datum $X_i$ is weighted according to its inverse sampling probability $1/G(X_i)$. The Efron-Petrosian NPMLE (Efron and Petrosian, 1999) $F_n$ can be derived in this way, since it corresponds to (\ref{Freverse}) with $F^*$ replaced by $F_n^*$ and $G$ replaced by its NPMLE $G_n$ (Shen, 2010). The asymptotic properties of $F_n$ and $G_n$ were recently detailed by de U\~na-\' Alvarez and Van Keilegom (2021), including uniform consistency and weak convergence results.

Unfortunately, $G_n$ \textcolor{black}{does not have a closed-form}; besides, the NPMLE $F_n$ may be non-unique, or even non-existing (Xiao and Hudgens, 2019). This motivates the interest in testing for the null hypothesis $\mathcal{H}_0: G(x)=\alpha$, $x \in \mathcal{S}_X$. Under $\mathcal{H}_0$ the ECDF $F_n^*$ is consistent for $F$, so there is no need to work with $F_n$; this is in agreement with the fact that, when $G$ is constant, there is no sampling bias on $X$, and the $X_i$'s are representative for the target population. Another way to write the null \textcolor{black}{hypothesis is} $\mathcal{H}_0: F(x)=F^*(x), x \in \mathcal{S}_X$; that both formulations are equivalent immediately follows from (\ref{Fstar}) or (\ref{Freverse}). \textcolor{black}{Certainly, by using (\ref{Fstar}), it is easy to see that the variance of $G(X)$ is zero whenever $F^*$ equals $F$ and, hence, $G(X)$ is constant almost surely in that case. The reverse implication is obviously true.} Therefore, it is natural to test for $\mathcal{H}_0$ by comparing $F_n$ to $F_n^*$. This is formalized in the following subsection.

\subsection{Test statistic}

A possible test statistic for $\mathcal{H}_0$ is given by the $L_\infty$ norm of $F_n-F_n^*$:

\begin{equation}
	D_n\equiv D_n(F_n,F_n^*)=\sup_{x\in \mathcal{S}_X}|F_n(x)-F_n^*(x)|.
	\label{Dn}
\end{equation}

\noindent Large values of $D_n$ lead to the rejection of the hypothesis of ignorable sampling bias; of course, here the vague meaning of 'large' must be formalized by using the null distribution of $D_n$, so the given significance level is respected. When (\ref{Dn}) leads to the acceptance of $\mathcal{H}_0$ one may say that there is no significant deviation between the Efron-Petrosian NPMLE $F_n$ and the ECDF $F_n^*$ or, alternatively, that the sampling bias is ignorable. In such a case, the simple estimator $F_n^*$ can be used to estimate the target $F$, and variance improvements with respect to $F_n$ can be obtained in this way. Moreover, under $\mathcal{H}_0$ the estimator $F_n^*$ is the NPMLE of $F$; this can be easily seen by decomposing the full likelihood (see (\ref{lik}) below) as a product of the marginal likelihood of the $X_i$'s and the conditional likelihood of the $(U_i,V_i)$'s given the $X_i$'s.

Since both $F_n^*$ and $F_n$ are concentrated on the $X_i$'s, a simple expression for (\ref{Dn}) can be derived. Specifically, the Efron-Petrosian NPMLE can be written as

\begin{equation}
	F_n(x)=\frac{\alpha_n}{n}\sum_{i=1}^n G_n(X_i)^{-1}I(X_i\leq x)
	\label{Fn}
\end{equation}

\noindent where $\alpha_n=\int G_ndF_n=n/\sum_{i=1}^n G_n(X_i)^{-1}$. Then,

\begin{equation}
	D_n=\max_{1\leq i\leq n}|F_n(X_i)-F_n^*(X_i)|=n^{-1}\alpha_n\max_{1\leq i\leq n}|\sum_{j=1}^n \left[\frac{1}{G_n(X_i)}-\frac{1}{\alpha_n}\right]I(X_j\leq X_i)|.
	\label{Dnmax}
\end{equation}

\noindent Equation (\ref{Dnmax}) indicates that $D_n$ is essentially the maximum cumulative difference between the inverse sampling probabilities $1/G_n(X_i)$ and $1/\alpha_n$. There exists no explicit formula for $G_n$; indeed, the couple $(F_n,G_n)$ is implicitly defined as the maximizer of the full likelihood of the $(X_i,U_i,V_i)$'s, which is given by

\begin{equation}
	\mathcal{L}_n = \left( \int \int_{u\leq x\leq v}dK(u,v) dF(x)\right)^{-n} \prod_{i=1}^n dF(X_i)dK(U_i,V_i).
	\label{lik}
\end{equation}

\noindent The assumed independence between $X$ and $(U,V)$ is critical to justify (\ref{lik}). Maximization of $\mathcal{L}_n$ with respect to both $F$ and $K$ leads to their NPMLEs, $F_n$ and $K_n$ respectively. Finally, the NPMLE of $G$ is computed from $K_n$ as $G_n(x)=\int_{u\leq x\leq v}dK_n(u,v)$.

The complicated structure of (\ref{lik}) results in a circularity, in the sense that $F_n$ depends on $G_n$, as indicated by (\ref{Fn}), while $G_n$ itself is depending on $F_n$. In practice, iterative algorithms that aim the maximization of $\mathcal{L}_n$ are used to compute the Efron-Petrosian NPMLE $F_n$ and/or the corresponding empirical sampling bias $G_n$. See de U\~na-\' Alvarez et al. (2021) for a thorough review of the NPMLE with doubly truncated data.

The limiting null distribution of $D_n$ is given in the following result. We will refer to the following conditions, where $[a_{U_1},b_{U_1}]$ and $[a_{V_1},b_{V_1}]$ denote the supports of $U_1$ and $V_1$:

\begin{itemize}
	\item [] (C1) $F$ has only a finite number of discontinuity points and is continuous everywhere else; $a_{U_1}<a_X<a_{V_1}$, $b_{U_1}<b_X<b_{V_1}$, $b_X-a_X<\infty$ and $P(V_1-U_1\geq \nu)=1$ for some $\nu>0$
	\item [] (C2) The marginal densities of $U_1$ and $V_1$ \textcolor{black}{have convex support and} are bounded on $\mathcal{S}_X$
\end{itemize}

Condition (C1) admits continuous and discrete distributions, provided that the number of point masses of $F$ is finite. Besides, this condition (C1), together with the convexity condition in (C2), implies that the sampling probability for $X$ is bounded away from zero, which is important in proofs. We mention that the inequalities $a_{U_1}<a_X<a_{V_1}$ and $b_{U_1}<b_X<b_{V_1}$ and the equality $P(V_1-U_1\geq \nu)=1$ in (C1) are only slightly stronger versions of the aforementioned identifiability conditions for $F$ and $K$. Finally, (C2) guarantees that the operator $\mathcal{A}$ appearing the asymptotic representation of $F_n$ is bounded; this is needed for \textcolor{black}{obtaining} the asymptotic properties of $F_n$. See de U\~na-\' Alvarez and Van Keilegom (2021) for further details and discussion.


\bigskip

THEOREM 1: \textit{Assume conditions (C1) and (C2). Then, under $\mathcal{H}_0$, it holds $\sqrt{n} D_n \rightarrow \sup_{x\in \mathcal{S}_X}|\mathcal{B}(x)|$ in distribution, where  $\mathcal{B}$ is a zero-mean Gaussian process.}

\bigskip

Theorem 1 follows from de U\~na-\' Alvarez and Van Keilegom (2021), Theorem 2.1, where an asymptotic representation for the centered Efron-Petrosian NPMLE $F_n(x)-F(x)$ as a sum of zero-mean iid random variables is given. Under $\mathcal{H}_0$ that representation still holds for $F_n(x)-F_n^*(x)$, although the extra term $F^*(x)-F_n^*(x)$ naturally appears. This however does not introduce new difficulties in proofs, due to the simple structure of the ECDF; note that the class of indicator functions is Donsker, so weak convergence is obtained. This, together with the continuous mapping theorem, is enough to conclude. The straightforward details are omitted.

The distribution of the limiting variable $\sup_{x\in \mathcal{S}_X}|\mathcal{B}(x)|$ in Theorem 1 is difficult to handle\textcolor{black}{; in general it will depend on an operator which does not have a closed-form}. To be specific, the process $\mathcal{B}(x)$ is the limit of $S_{1n}(x)+S_{2n}(x)$, $x \in \mathcal{S}_X$, where $S_{1n}(x)=n^{-1/2}\sum_{i=1}^n \mathcal{A}[h_i](x)$ for a certain linear operator $\mathcal{A}$ and certain iid random functions $h_i$, and $S_{2n}(x)=-n^{-1/2}\sum_{i=1}^n (I(X_i\leq x)-F(x))$. \textcolor{black}{\textcolor{black}{Moreover}, $\mathcal{A}$ equals the sum of an infinite series, $\sum_{r=0}^\infty A^r$, where $A$ is an operator depending on population parameters in a complicated way (de U\~na-\'Alvarez and Van Keilegom, 2021). Thus, it is unclear how the asymptotic law arising from the term $S_{1n}(x)+S_{2n}(x)$ can be used for practical purposes.} Interestingly, both $S_{1n}(x)$ and $S_{2n}(x)$ are zero-mean under the null hypothesis of ignorable sampling bias; however, in general the expectation of $S_{2n}(x)$ is given by $n^{1/2}(F(x)-F^*(x))$, which does not vanish when $\mathcal{H}_0$ is false. This guarantees the consistency of $D_n$ as the sample size $n$ grows.

Due to the aforementioned complexity of the asymptotic null distribution of $D_n$, a bootstrap approximation is proposed in the following subsection.

\subsection{Bootstrap approximation}

Bootstrap methods for randomly truncated data have been widely investigated in the literature. \textcolor{black}{In the left-truncated setting, Gross and Lai (1996) discussed and compared two different bootstrap resampling plans: the simple bootstrap and the obvious bootstrap.} For doubly truncated data, the simple and the obvious bootstrap were used in Moreira and de U\~na-\' Alvarez (2010) to approximate the sampling distribution of the Efron-Petrosian NPMLE $F_n$. The simple bootstrap resamples from the $(X_i,U_i,V_i)$'s with replacement. On \textcolor{black}{the other hand}, the obvious boostrap independently resamples the target variable $X$ and the truncating couple $(U,V)$ from $F_n$ and $K_n$ respectively; then, an acceptance/rejection rule is implemented, so only the triplets satisfying $U\leq X\leq V$ are retained. \textcolor{black}{As discussed in the aforementioned works, these two bootstraps are not equivalent. In particular, with the obvious bootstrap triplets $(X_i,U_j,V_j)$, $i \neq j$, are possible as long as $U_j\leq X_i\leq V_j$ is satisfied, while only triplets belonging to the original sample are allowed by the simple bootstrap. This makes a difference with respect to the right-censored setting, in which the simple bootstrap and the obvious bootstrap induce the same probability law (Efron, 1981).}

\textcolor{black}{In the testing framework, \textcolor{black}{bootstrap methods aim the estimation of the null distribution of the test statistic. For this, the bootstrap usually incorporates the null hypothesis at some stage of the resampling algorithm. Unfortunately, it is unclear how this can be done in the setting of this paper, since the null hypothesis does not determine the distribution of the random variables $X$ and $(U,V)$.} A possibility is to apply the obvious bootstrap described above but with $F_n$ replaced by $F_n^*$ when resampling the target variable, so the hypothesis $F=F^*$ is mimicked. Preliminary simulations indicate however that such bootstrap approximation may be conservative (results not shown). An alternative route is to modify the structure of the bootstrap test statistic in such a way that the simple bootstrap becomes applicable; see Mart\'inez-Camblor and Corral (2012) for a deeper insight. We will follow this approach in order to propose a bootstrap approximation for the null distribution of $D_n$. }

\textcolor{black}{Introduce $D_n^{(1)}=\sup_{x \in \mathcal{S}_X}|F_n(x)-F(x)+F^*(x)-F^*_n(x)|$. Note that equality $D_n=D_n^{(1)}$ holds under $\mathcal{H}_0$, while $D_n \neq D_n^{(1)}$ under the alternative.} Given the bootstrap triplets $(X_i^b,U_i^b,V_i^b)$, $1\leq i\leq n$, sampled with replacement from the $(X_i,U_i,V_i)$'s, define $D_n^b=\sup_{x \in \mathcal{S}_X}|F_n^b(x)-F_n(x)+F_n^*(x)-F^{*,b}_n(x)|$. Here, $F_n^b$ and $F_n^{*,b}$ denote the estimators $F_n$ and $F_n^*$ when computed from the bootstrap resample. \textcolor{black}{For left-truncated data, Gross and Lai (1996) established under certain conditions the validity of the simple bootstrap to approximate the distribution of the NPMLE $F_n$, a result that can be readily extended to $F_n(x)-F_n^*(x)$. This proves that the bootstrap process $F_n^b(x)-F_n(x)+F_n^*(x)-F^{*,b}_n(x)$, $x\in \mathcal{S}_X$, consistently approximates the distribution of $D_n^{(1)}$ in the left-truncated setting. The formal extension of such result to double truncation is not immediate, however. For instance, Edgeworth expansions as those invoked by Gross and Lai (1996) \textcolor{black}{have not been developed} for the Efron-Petrosian estimator; the complex nature of $F_n$ discussed in Section 2.2 makes this difficult. Importantly, simulation results in the following section, see also the supporting information, reveal that the proposed approximation works well when testing for $\mathcal{H}_0$; this suggests that the simple bootstrap is consistent under double truncation too. On the other hand, our results are in agreement with Moreira and de U\~na-\' Alvarez (2010), who showed that the simple bootstrap performs satisfactorily in the construction of pointwise confidence intervals for a doubly truncated CDF.} \textcolor{black}{Note that the bootstrap approximation proposed for the test statistic $D_n$ does not draw resamples under the null hypothesis; rather, the bootstrap version of the test statistic $D_n^b$ is centered around $F_n(x)-F_n^*(x)$. This moves the testing problem close to the confidence interval setting.}


In practice, $D_n^b$ can be implemented as a maximum along the $X_i$'s, since (with the simple bootstrap) the $X_i^b$'s are by force a subset of the original data. We propose to compute $D_n^b$, as defined here, for a large number of bootstrap replicates $B$; then, the null distribution of $D_n$ is approximated from its bootstrap evaluations $D_n^b$, $1\leq b\leq B$. Equivalently, \textcolor{black}{the} bootstrap \textit{P}-value is computed as

\begin{equation}
	p^B=B^{-1}\sum_{b=1}^B I(D_n^b \geq D_n).
	\label{eq:boot_p}
\end{equation}

\noindent The null hypothesis of ignorable sampling bias is rejected when $p^B$ is smaller than or equal to the nominal significance level for the test. \textcolor{black}{Note that, since the distribution of the $D_n^b$'s approximate that of $D_n^{(1)}$, and since $D_n=D_n^{(1)}$ under $\mathcal{H}_0$, the P-value in (\ref{eq:boot_p}) \textcolor{black}{is conjectured to follow} a uniform distribution under the null hypothesis. On the other hand, in general the distribution of $D_n$ is shifted to the right when compared to $D_n^{(1)}$, the shift increasing as $F^*$ departs from $F$. \textcolor{black}{Therefore, a natural conjecture is} that the proposed P-value is stochastically dominated by a uniform random variable under the alternative, anticipating the consistency of the proposed bootstrap approach.}

The performance of the test statistic $D_n$ with the given bootstrap approximation is investigated by simulations in the following section.

\section{Simulation study}

In this section we investigate the finite sample performance of the test statistic $D_n$ through simulations. Two different models are considered: the first one corresponds to interval sampling (Model 1), while in the second one the truncation limits are simulated independently (Model 2). Specifically, for a target variable $X$ supported on the unit interval $\mathcal{S}_X=[0,1]$ and certain parameter values $\varsigma>0$ and $\rho>0$ we consider the following scenarios:

\begin{itemize}
	\item [] M1 (Model 1). $U=(1+\varsigma)Z^\rho-\varsigma$ and $V=U+\varsigma$, where $Z \sim U(0,1)$ is independent of $X$;
	\item [] M2 (Model 2). $U=(1+\varsigma)Z_1-\varsigma$ and $V=\varsigma(Z_2^{-\rho}-1)$, where $Z_i \sim U(0,1)$, $i=1,2$, are independent random variables and independent of $X$
\end{itemize}

In Model 1 (interval sampling) the $\varsigma$ parameter stands for the width of the sampling interval. In this Model 1, the left-truncation limit $U$ is supported on $(-\varsigma,1)$, while $V$ is supported on $(0,1+\varsigma)$; this implies that both $F$ and $K$ are identifiable. In Model 2 (independent truncation limits), $U$ is again supported on $(-\varsigma,1)$ but $V$ is now supported on $(0,\infty)$. In this case, $F$ is identifiable, but $K$ is not; this is because $V<U$ may occur (the couple $(U,V)$ should be redefined as conditionally on $U\leq V$ if the identifiability of $K$ is aimed). In both models M1 and M2 the sampling bias is ignorable only when $\rho=1$. Specifically, for Model 1 and $x \in (0,1)$ it holds

\begin{equation}
	G(x)=(1+\varsigma)^{-1/\rho}[(x+\varsigma)^{1/\rho}-x^{1/\rho}],
	\label{eq:G_M1}
\end{equation}

\noindent so when $\rho=1$ the sampling probability $G(x)$ is free of $X$ and the truncation proportion is $(1+\varsigma)^{-1}$. On the other hand, for Model 2 one gets

\begin{equation}
	G(x)=\varsigma^{1/\rho}(1+\varsigma)^{-1}(x+\varsigma)^{1-1/\rho};
	\label{eq:G_M2}
\end{equation}

\noindent again, (\ref{eq:G_M2}) is constant when $\rho=1$, giving the same truncation rate as in Model 1. Smaller values of $\varsigma$ result in a larger proportion of truncated data. For instance, in Model 1 the truncation rate for $\rho=1$, $\rho=2$ and $\rho=6$ is 50\%, 61\% and 81\% ($\varsigma=1$), or 67\%, 74\% and 87\% ($\varsigma=1/2$). For Model 2, the truncation rate is 39\% ($\rho=2,\varsigma=1$), 30\% ($\rho=6,\varsigma=1$), 53\% ($\rho=2,\varsigma=1/2$) or 41\% ($\rho=6,\varsigma=1/2$). We will investigate the influence of the $\varsigma$ parameter in the performance of the test; specifically, values $\varsigma=1$ and $\varsigma=1/2$ will be considered.



\begin{figure}[!p]
	\centering\includegraphics[width=1\textwidth]{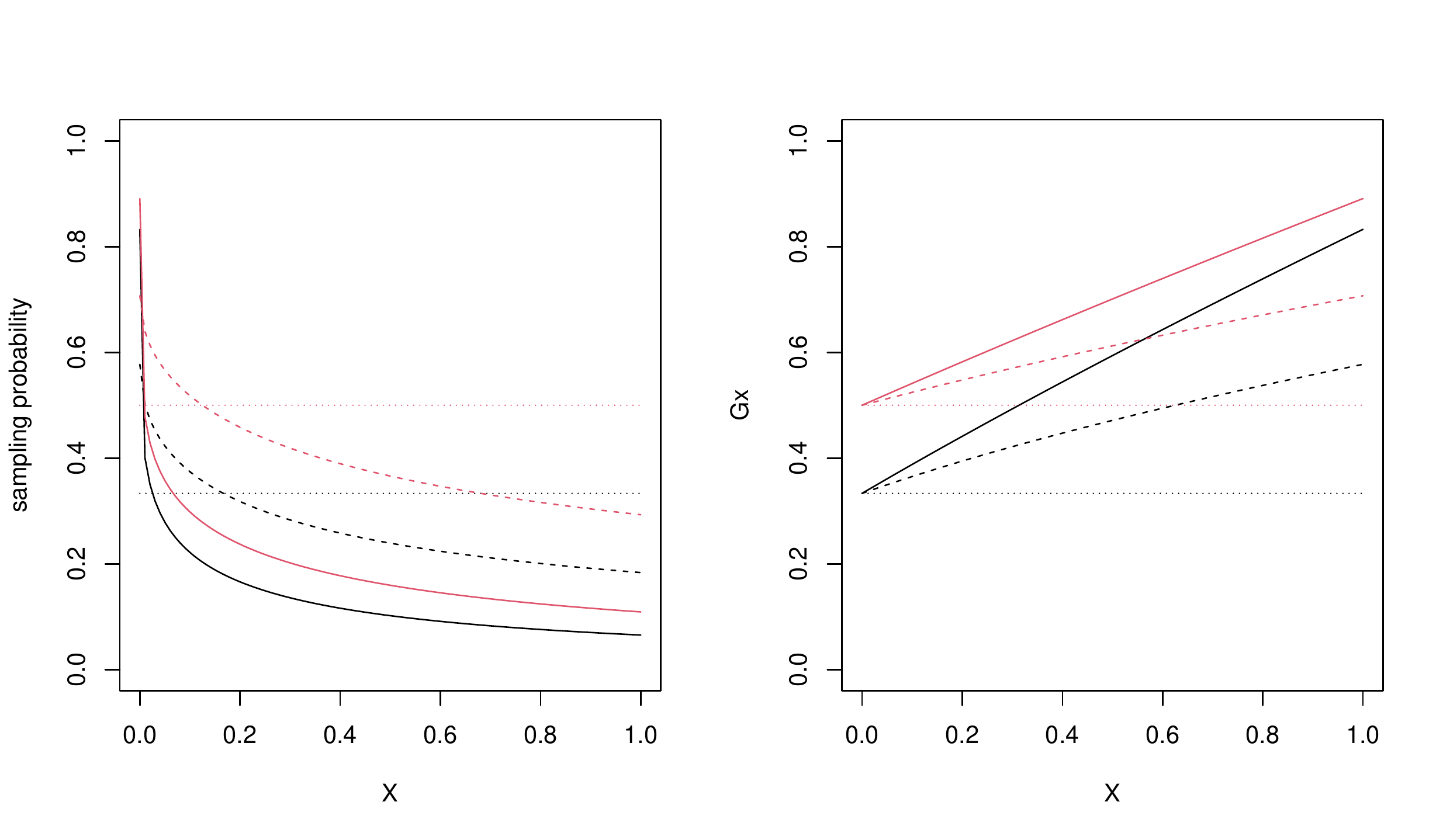}
	\caption{Sampling bias for simulated Model 1 (left) and Model 2 (right): $\rho=1$ (dotted line), $\rho=2$ (dashed line) and $\rho=6$ (solid line). Black and red lines correspond to $\varsigma=1/2$ and $\varsigma=1$ respectively.}
	\label{Fig1}
\end{figure}

Besides the null case $\rho=1$ we will consider weak ($\rho=2$) and strong ($\rho=6$) deviations from the hypothesis of ignorable sampling bias under Model 1 and Model 2. The sampling probabilities (\ref{eq:G_M1}) and (\ref{eq:G_M2}) for the several choices of $\rho$ and $\varsigma$ are depicted in Figure \ref{Fig1}. From this Figure \ref{Fig1} it is seen that, when $\rho \neq 1$, the sampling probability decreases (Model 1) or increases (Model 2) as $X$ grows, with a more clear violation of $\mathcal{H}_0$ when $\rho=6$. Finally, the target variable $X$ will be taken as uniformly distributed on the unit interval or, alternatively, distributed as a $Beta(a,b)$ model with shape parameters $(a,b)=(1,1/2)$ and $(a,b)=(1/2,1)$. This will allow for the investigation of the sensitiveness of $D_n$ to changes in $F$.

In Table \ref{Table:M} the proportion of rejections of $\mathcal{H}_0$ performed by $D_n$ along \textcolor{black}{1000} Monte Carlo replicates is given; the results corresponds to a uniformly distributed target $X$. We considered sample sizes $n \in \{100,200\}$ and nominal significance levels $\gamma \in \{0.1,0.05,0.01\}$. The number of bootstrap replicates was \textcolor{black}{$B=500$}. From Table \ref{Table:M} it is seen that the proposed test respects the nominal level fairly well. On the other hand, the power of the test increases with the sample size and the degree of departure from the hypothesis of ignorable sampling bias, as expected. Finally, the influence of $\varsigma$ is more subtle. For Model 1, the statistical power increases with $\varsigma$, and this is well connected to the meaning of this parameter, which is the width of the sampling interval under M1. For Model 2, the opposite occurs. This can be explained from the fact that, when $\varsigma=1/2$, the sampling probability $G$ under M2 departs more from a constant than with $\varsigma=1$ (Figure \ref{Fig1}, right).\textcolor{black}{ Violations of the level occurred in some of the scenarios when using a smaller number of bootstrap replicates ($B=300$, results not shown); this sensitivity should be taken into account when using the proposed bootstrap approximation in practice.}

\begin{table}
	\caption{Proportion of rejections of $\mathcal{H}_0$ performed by $D_n$ along \textcolor{black}{1000} Monte Carlo trials, Model 1 (interval sampling) and Model 2 (independent truncation limits). The nominal level of the test is $\gamma$. The null hypothesis corresponds to $\rho=1$, while $\rho=2$ and $\rho=6$ represent weak and strong deviation from $\mathcal{H}_0$, respectively. \textcolor{black}{The number of trials that were discarded for $\rho=(1,2,6)$ because the Efron-Petrosian NPMLE did not exist or was not unique is indicated between brackets at the right of the $\varsigma$ value}.\textcolor{black}{The number of bootstrap replicates was 500.}
		\label{Table:M}}
	\begin{center}
		\begin{tabular}{rrrrrrrrrr}
			& & $\rho=1$ & & & $\rho=2$ & & & $\rho=6$ & \\ 
			$\gamma$: & $0.1$ & $0.05$ & $0.01$ & $0.1$ & $0.05$ & $0.01$ & $0.1$ & $0.05$ & $0.01$\\\hline
			& & & & & & & & &\\
			Model 1, $n=100$  & & & & & & & & & \\
			$\varsigma=1$ $(18,49,118)$ & .0866 & .0346 & .0031 & .5342 & .4038 & .1672 & .9524 & .8980 & .6474 \\
			$\varsigma=1/2$ $(36,63,153)$ & .0788 & .0342 & .0062 & .2241 & .1249 & .0395 & .5502 & .3601 & .0874 \\
			& & & & & & & & & \\
			Model 1, $n=200$ & & & & & & & & & \\
			$\varsigma=1$ $(6,25,79)$ & .0755 & .0493 & .0131 & .8585 & .7713 & .5149 & .9957 & .9946 & .9739 \\
			$\varsigma=1/2$ $(18,49,97)$ & .0876 & .0407 & .0051 & .4585 & .3291 & .1367 & .8959 & .7697 & .4363 \\
			& & & & & & & & &\\
			Model 2, $n=100$  & & & & && & & & \\
			$\varsigma=1$ $(18,18,6)$ & .0804 & .0397 & .0071 & .5193 & .4012 & .1670 & .9789 & .9447 & .7736\\
			$\varsigma=1/2$ $(33,23,16)$ & .0889 & .0465 & .0062 & .6888 & .5333 &  .2651 & .9990 & .9878 & .9258\\
			& & & & & & & & & \\
			Model 2, $n=200$ & & & & & & & & & \\
			$\varsigma=1$ $(7,9,4)$ & .0886 & .0483 & .0030 & .8214 & .7215 & .4480 & 1.000 & 1.000 & .9970 \\
			$\varsigma=1/2$ $(16,8,5)$ & .0884 & .0467 & .0071 & .9234 & .8569 & .6522 & 1.000 & 1.000 & 1.000 \\ 
			& & & & & & & & &\\
			\hline
		\end{tabular}
	\end{center}
\end{table}

The simulations were repeated with a different distribution for the target variable $X$. As mentioned, $Beta(1,1/2)$ and $Beta(1/2,1)$ models were considered to this end. The results\textcolor{black}{, see Supplementary Tables 1 and 2 in the supporting information,} went in the expected direction; that is, the power of $D_n$ decreased as the target distribution concentrated in areas along which $G$ was relatively flat. \textcolor{black}{To be specific,} the power of $D_n$ for $X\sim Beta(1,1/2)$ (which shifts the uniform density to the right) was smaller compared to the figures in Table \ref{Table:M}. Naturally, the situation was just the opposite with $X\sim Beta(1/2,1)$. \textcolor{black}{This is in agreement with the fact that, for both Model 1 and Model 2, the absolute value of the first derivative of $G(x)$ is monotone decreasing.}

In the simulation study an issue related to the possible non-existence or non-uniqueness of the Efron-Petrosian NPMLE occurred. Specifically, for a small number of trials the estimator $F_n$ could not be computed in a reliable way. These trials were eliminated when computing the empirical rejection levels attached to $D_n$. \textcolor{black}{In Table \ref{Table:M} the exact number of discarded samples is provided; it is seen that the problem occurs less frequently when increasing the sample size.} The same issue appeared in the bootstrap resamples. In this case, the bootstrap \textit{P}-value $p^B$ in (\ref{eq:boot_p}) was computed from the available evaluations of $D_n^b$; anyway, these were the total amount of \textcolor{black}{$B=500$} evaluations for most of the simulated trials. \textcolor{black}{In practice, potential issues with the computation of the NPMLE can be avoided through the preliminary fitting of a parametric truncation model; of course, this may induce some estimation bias, particularly when the parametric family is miss-specified. \textcolor{black}{To be explicit, in the setting of parametric truncation the weights $G_n(X_i)$ in (\ref{Fn}) are replaced by $G(X_i;\hat \theta)$, where $G(x;\hat \theta)=\int_{u\leq x\leq v}dK(u,v;\hat \theta)$ is the sampling probability for $X=x$ under a given parametric family $K(\cdot,\cdot;\theta)$ for the truncation CDF. This leads to an alternative estimator for $F$, of semiparametric nature. The $\theta$ parameter can be estimated by maximizing the conditional likelihood of the $(U_i,V_i)$'s given the $X_i$'s. See Moreira et al. (2014) for more on this. Indeed, the parametric truncation setup provides an alternative test for ignorable sampling bias because, when $G(\cdot;\theta_0)$ is flat for some $\theta_0$ in the parametric space, a test for the null hypothesis $\theta=\theta_0$ is valid for that aim. A limitation of this alternative approach is that it is not omnibus, since it may fail to detect departures from the null hypothesis when the parametric family is miss-specified.}}

\section{Real data applications}
\label{sec3}

\subsection{Childhood cancer data}

We consider all the children diagnosed from cancer in the region of North Portugal (which includes the districts of Porto, Braga, Bragan\c{c}a, Vila Real and Viana do Castelo) between January 1999 and December 2003 (Moreira and de U\~na-\'Alvarez, 2010).  The number of cases was 409. The target variable is the age at diagnosis $X$ (in days) and, thus, it is doubly truncated due to the interval sampling. The truncating values $(U,V)$ are determined by the birth dates, and it holds $V=U+1825$ (interval width of 5 years). Information on $X$ was missing for three cases; hence, the sample size is $n=406$. The data are available within the data frame \texttt{ChildCancer} of the \texttt{R} package \texttt{DTDA}.

The value of the test statistic $D_n$ was $0.0206$, with corresponding \textit{P}-value $0.9120$ based on $B=500$ bootstrap resamples. Thus the test largely accepts the null hypothesis of ignorable sampling bias. This is in agreement with the informal analysis of Moreira and de U\~na-\'Alvarez (2010), who obtained a truncation distribution close to uniform for this dataset.

\begin{figure}[!p]
	\centering\includegraphics[width=1\textwidth]{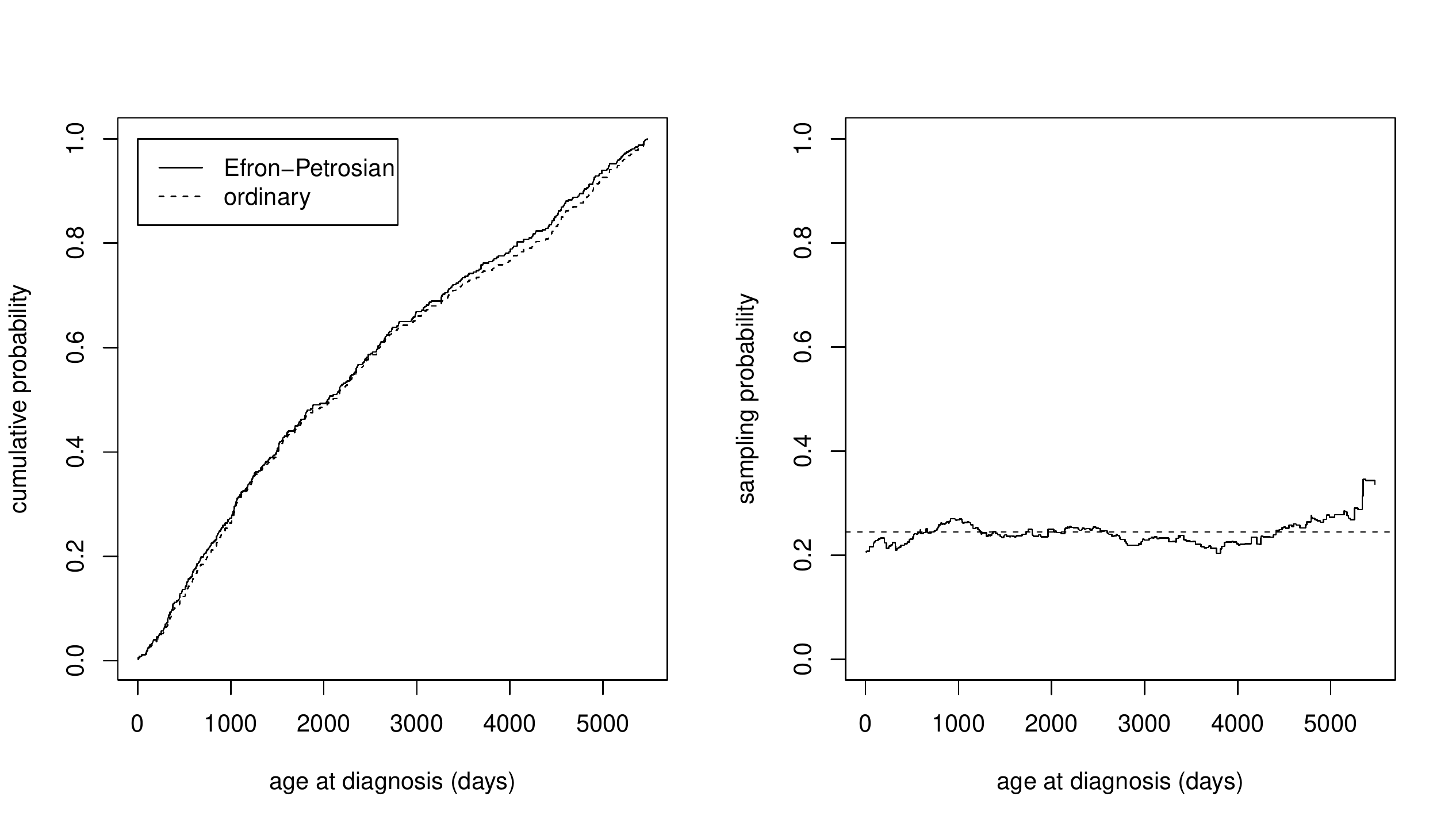}
	\caption{Left: Efron-Petrosian NPMLE (solid line) and ECDF (dashed line). Right: NPMLE of the sampling probability (solid line) and sampling probability under the hypothesis of ignorable sampling bias (dashed line). Childhood cancer data.}
	\label{Fig2}
\end{figure}

In Figure \ref{Fig2} the goodness-of-fit plots for $\mathcal{H}_0$ corresponding to the processes $F_n(x)-F_n^*(x)$ (left) and $G_n(x)-\alpha_n$ (right) are provided; the estimated proportion of truncation is, in this case, $1-\alpha_n=0.7557$. From this Figure \ref{Fig2} it is seen that the ECDF is close to the Efron-Petrosian estimator, and that the sampling probability is roughly constant. The fact that an ignorable sampling bias holds in this case can be explained from the homogeneity of the birth process for the individuals who will develop cancer along their childhood.

\begin{figure}[!p]
	\centering\includegraphics[width=0.65\textwidth]{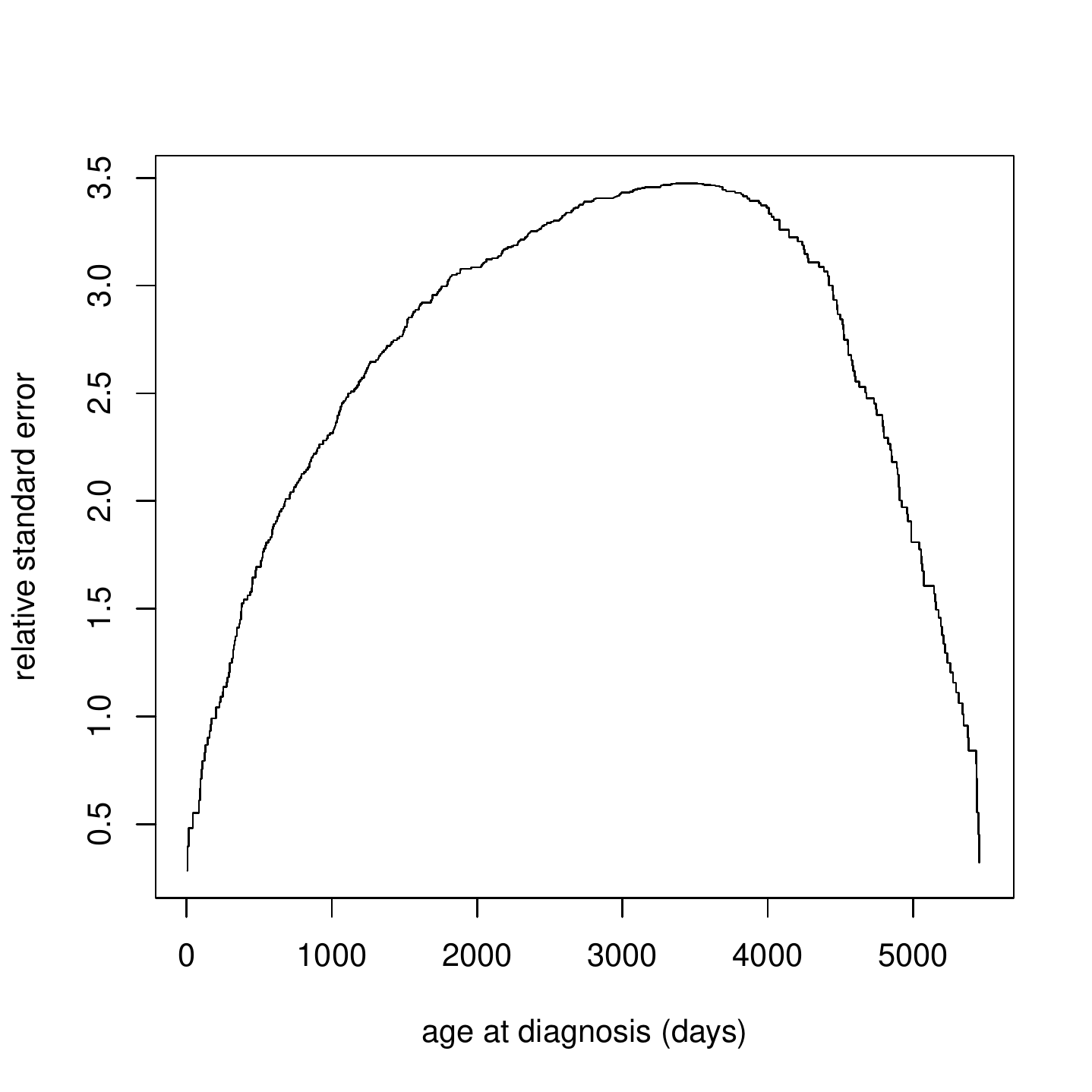}
	\caption{Standard error of the Efron-Petrosian NPMLE relative to the ordinary empirical cumulative distribution function. Childhood cancer data.}
	\label{Fig:rse_CCD}
\end{figure}

Variance improvements when using $F_n^*(x)$ instead of $F_n(x)$ to analyze the data on childhood cancer are depicted in Figure \ref{Fig:rse_CCD}. Specifically, the ratio $\sigma_n(x) / \sigma_n^*(x)$ for $x \in \{X_i,1\leq i\leq n\}$ is displayed, where $\sigma_n^*(x)=(F_n^*(x)(1-F_n^*(x))/n)^{1/2}$ is the usual empirical standard error of $F_n^*(x)$ and $\sigma_n(x)$ is the standard error of the Efron-Petrosian estimator based on the simple bootstrap ($B=500$ replicates). From Figure \ref{Fig:rse_CCD} it is seen that $\sigma_n(x)$ is several times larger than $\sigma^*_n(x)$ in most of the support of $X$; interestingly, the standard error of $F_n(x)$ relative to $F_n^*(x)$ may be as large as $3.5$ at particular quantiles $x$. This illustrates how testing for $\mathcal{H}_0$ may help to reduce the estimation variance and, hence, to facilitate the statistical inference from the doubly truncated outcomes.

\subsection{Parkinson's disease data}

Clark et al. (2011) studied the association between genetic information and age of onset of Parkinson's disease. In that study, in order to eliminate potential biases coming from different survival profiles, the selected patients were those with the DNA sample taken eight years at maximum after the onset of Parkinson. Therefore, the age at onset $X$ is truncated from the right by the age at blood sampling $V$, and left-truncated by $U=V-8$ (age in years). Two different groups of patients were considered: early onset, with ages at onset ranging between 35 and 55 years ($n=99$); and late onset, for which the ages range between 63 and 87 years ($n=100$). These two datasets are available from \texttt{PDearly} and \texttt{PDlate} objects in the package \texttt{DTDA}. Truncation values are missing for two patients in the early onset group, and these cases were removed for our analyses; the final sample size in this group is thus $n=97$.

\begin{figure}[!p]
	\centering\includegraphics[width=1\textwidth]{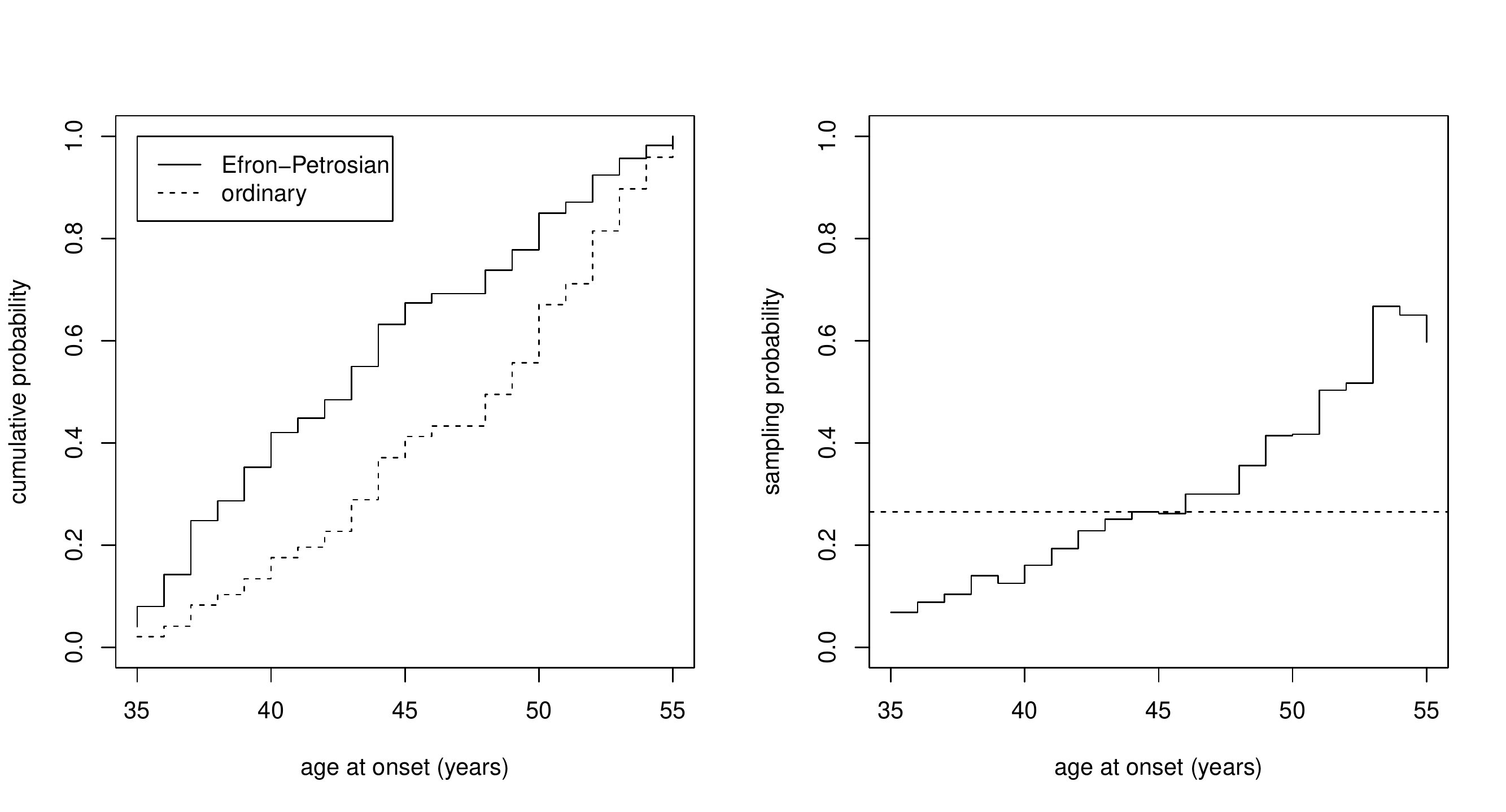}
	\caption{Left: Efron-Petrosian NPMLE (solid line) and ECDF (dashed line). Right: NPMLE of the sampling probability (solid line) and sampling probability under the hypothesis of ignorable sampling bias (dashed line). Parkinson's disease data, early onset.}
	\label{Fig:FG_PDearly}
\end{figure}

\begin{figure}[!p]
	\centering\includegraphics[width=1\textwidth]{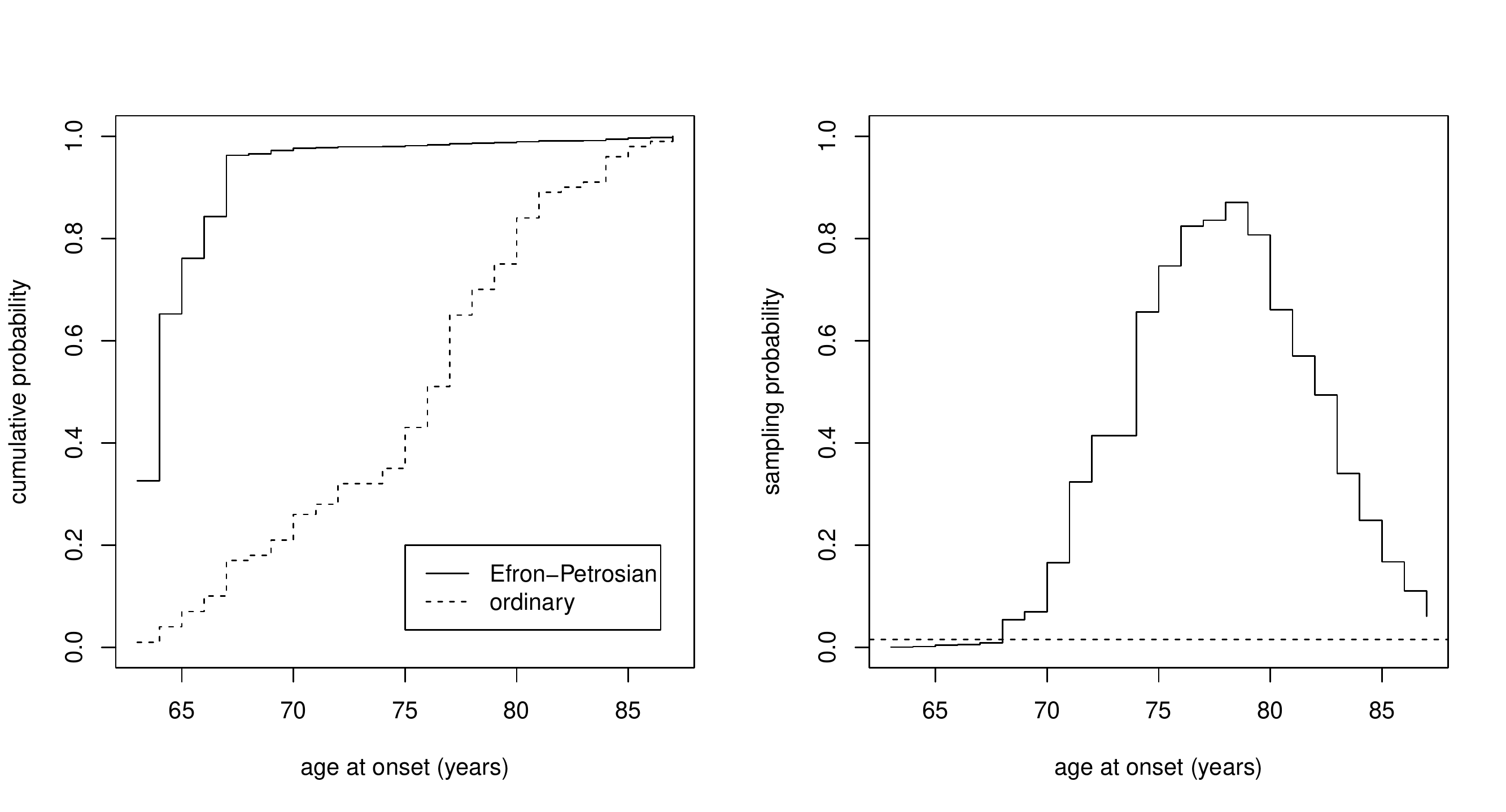}
	\caption{Left: Efron-Petrosian NPMLE (solid line) and ECDF (dashed line). Right: NPMLE of the sampling probability (solid line) and sampling probability under the hypothesis of ignorable sampling bias (dashed line). Parkinson's disease data, late onset.}
	\label{Fig:FG_PDlate}
\end{figure}

The informal graphical assessment for ignorable sampling bias for the early and late onset groups is given in Figures \ref{Fig:FG_PDearly} and \ref{Fig:FG_PDlate} respectively. In both cases \textcolor{black}{a sampling} bias is revealed; this is much clearer for the late onset group, in which the sampling probabilities are extremely low for ages below $69$ years (Figure \ref{Fig:FG_PDlate}, right). This results in a clear departure between $F_n$ and $F_n^*$ at the left tail (Figure \ref{Fig:FG_PDlate}, left). The evidences against $\mathcal{H}_0$ in the early onset group are weaker, although the empirical sampling probability $G_n$ exhibits a clearly increasing shape. The formal testing of $\mathcal{H}_0$ through $D_n$ gave the following results (\textit{P}-values $p$ computed from 500 replicates; 30 resamples were removed for the late onset group due to the non-existence/non-uniqueness of the NPMLE): $D_n=0.2612$ and $p=0.0520$ (early onset), and $D_n=0.7929$, $p=0.0022$ (late onset). Therefore, at significance level $0.05$ the test accepts the null for the early onset group, but it rejects it for the late onset. 

For completeness, the accuracy of $F_n$ relative to $F_n^*$ in both groups was calculated (results not shown), even when one would probably be in favour of using the Efron-Petrosian NPMLE for the Parkinson's disease study, according to the attained \textit{P}-values. Similarly as in the childhood cancer study, the standard error of the Efron-Petrosian estimator was several times that of $F_n^*(x)$ for most of the $x$-values (the $X_i$'s), with a larger relative deficiency of $F_n$ at the left tail of the distribution. The situation for the late onset group was critical, in the sense that the standard error of $F_n$ was more than ten times larger at specific quantiles. Unfortunately, the formal tests performed by $D_n$ give few chances to work with $F_n^*$ in this case.


\section{Discussion}
\label{sec4}

Random truncation induces most of the times a sampling bias on the target variable. This is always the case with left- or right-truncation, often encountered in the analysis of time-to-event data, where proper corrections are needed. However, with double truncation the situation may be different, since the truncation limits may compensate each other so the sampling bias becomes negligible. If that is the case, ordinary statistical procedures are applicable, thus simplifying the estimation and inference.

In this paper a formal test for the null hypothesis of ignorable sampling bias under random double truncation has been proposed. The test is based on the maximum departure between the Efron-Petrosian NPMLE and the ECDF. The asymptotic null distribution of the test statistic has been established, and a bootstrap procedure for the practical application of the test has been designed. Simulation studies have been conducted in order to investigate the finite sample performance of the test. The method has been found to respect the nominal level well, while exhibiting a power that increases with the sample size and the degree of violation of the null hypothesis. Applications to data on childhood cancer and Parkinson's disease have served to further illustrate the proposed method, including the variance improvements entailed by the acceptance of $\mathcal{H}_0$ when estimating the target distribution $F$.

Test statistics for $\mathcal{H}_0$ based on the sampling probability process $G_n(x)-\alpha_n$, $x \in \mathcal{S}_X$, could be considered too. Preliminary simulations performed by the author (results not shown) indicate that the test based on the supremum norm of this alternative process, $D_n^G$ say, does not dominate (nor is dominated by) $D_n$ in the sense of the power. Importantly, since the jump points of $G_n(x)$ correspond to the truncation values, the practical implementation and interpretation of $D_n^G$ requires some care. It is also possible to consider distances other than the supremum (Kolmogorov-Smirnov type) norm to redefine the test statistic $D_n$, such as Cramér-von Mises or Anderson-Darling type distances. This is currently under study and the corresponding results will be presented elsewhere.

Another possible route to explore when looking for powerful testing procedures is that given by smooth tests. With smooth tests density functions, rather than cumulative distributions, are compared. This may result in power improvements when the bandwidth factor is properly chosen; see for instance Mart\'inez-Camblor and de U\~na-\' Alvarez (2009). Recently, tests based on the comparison of empirical characteristic functions have been investigated in a variety of settings (Cousido-Rocha et al., 2019; Henze and Jim\' enez-Gamero, 2021). Such approach could be brought here too in order to construct a test for $\mathcal{H}_0$.


The null hypothesis of ignorable sampling bias can be written as $\mathcal{H}_0:a(x)=1$, $x \in \mathcal{S}_X$, where $a(x)=\alpha^{-1}G(x)$ is the normalized sampling probability. A generalization of this testing problem is the one in which the null states $a(x)=a_0(x)$, $x\in \mathcal{S}_X$, for a fully specified function $a_0(x)$. This is relevant when there exists information on the sampling bias other than ignorability; for instance, with interval sampling such information could be given by general population registries reporting birth rates for the process of interest. Under this generalized null hypothesis, the NPMLE of $F$ is just the inverse-probability-weighted estimator, $F_n^0$ say, which attaches weight $a_0(X_i)^{-1}$ to $X_i$, $1\leq i\leq n$. Obviously, $D_n$ can be generalized for this problem, becoming the maximum deviation between the $F_n(X_i)$'s and the $F_n^0(X_i)$'s. Formal theory can be derived similarly as in Theorem 1, although regularity conditions on the function $a_0(x)$ must be imposed. When the fully specified function $a_0(x)$ in $\mathcal{H}_0$ is replaced by a parametric family things are more complicated; the fact that the function $a(x)$ does not characterize the truncation distribution is responsible for this. Minimum-distance and pseudo-likelihood approaches are possible; the practical performance of such estimators and the development of the corresponding asymptotic theory are interesting topics for our future research.



\section*{Acknowledgements}
Work supported by the Grant PID2020-118101GB-I00, Ministerio de Ciencia e Innovaci\' on.





\section*{Supporting Information}

\textcolor{black}{Supplementary Tables 1 and 2 referenced in Section 3 are available from the author upon request. Same applies to the code to reproduce the simulation results in Section 3 and the real data analyses in Section 4.}

\section*{References}

\begin{itemize}
	\item [] Clark, J., Reddy, S. Zheng, K., Betensky, R.A. and Simon, D.K. (2011).
	Association
	of {PGC}-1alphapolymorphisms with age of onset and risk of {P}arkinson's disease.
	{\em BMC Medical Genetics} {\bfseries 12}, 69.
	\item [] Cousido-Rocha, M., {de U\~{n}a-\'{A}lvarez}, J. and Hart, J. (2019).
	A two-sample test for the 
	equality of univariate marginal distributions for high-dimensional data.
	{\em Journal of Multivariate Analysis} {\bfseries 174}, 104537.
	\item [] {de U\~{n}a-\'{A}lvarez}, J. (2020a).
	R packages for the statistical analysis of doubly truncated data: a review.
	{\em arXiv} {\bfseries 2004.08978}, 1--23.
	\item [] {de U\~{n}a-\'{A}lvarez}, J. (2020b).
	Nonparametric estimation of the cumulative incidences 
	of competing risks under double truncation.
	{\em Biometrical Journal} {\bfseries 62}, 852--867.
	\item [] {de U\~na-\'Alvarez}, J., Moreira, C. and Crujeiras, R.M. (2021).
	{\em The Statistical Analysis of Doubly Truncated Data: With Applications in R}.
	Hoboken, NJ: John Wiley.
	\item [] {de U\~{n}a-\'{A}lvarez}, J. and {Van Keilegom}, I. (2021).
	Efron-{P}etrosian integrals for doubly truncated data with covariates: an asymptotic analysis.
	{\em Bernoulli} {\bfseries 27}, 249--273.
	\item [] Efron, B. (1981). Censored data and the bootstrap. {\em Journal of the American Statistical Association} {\bfseries 76}, 312--319.
	\item [] Efron, B. and Petrosian, V. (1999).
	Nonparametric methods for doubly truncated data.
	{\em Journal of the American Statistical Association} {\bfseries 94}, 824--834.
	\item [] Emura, T., Hu, Y.H. and Huang, C.Y. (2020).
	\texttt{double.truncation}: Analysis of Doubly Truncated Data.
	{\em \texttt{R} package version 1.7}, https://CRAN.R-project.org/package=double.truncation.
	\item [] Emura, T., Hu, Y.H. and Konno, Y. (2017).
	Asymptotic inference for maximum likelihood estimators under the special exponential family with double‐truncation.
	{\em Statistical Papers} {\bfseries 58}, 877--909.
	\item [] Emura, T., Konno, Y. and Michimae, H. (2015).
	Statistical inference based on the nonparametric maximum likelihood estimator under double‐truncation.
	{\em Lifetime Data Analysis} {\bfseries 21}, 397--418.
	\item [] Gross, S.T. and Lai, T.L. (1996).
	Bootstrap methods for truncated and censored data.
	{\em Statistica Sinica} {\bfseries 6}, 509--530.
	\item [] Henze, N. and Jim\' enez-Gamero, M.D. (2021).
	A test for {G}aussianity in {H}ilbert spaces via the
	empirical characteristic functional.
	{\em Scandinavian Journal of Statistics} {\bfseries 48}, 406--428.
	\item [] Martin, E.C. and Betensky, R.A. (2005).
	Testing quasi-independence of failure and truncation times via conditional Kendall's tau.
	{\em Journal of the American Statistical Association} {\bfseries 100}, 484--492.
	\item [] Mandel, M., {de U\~{n}a-\'{A}lvarez}, J., Simon, D.K. and Betensky, R.A. (2018).
	Inverse probability weighted {Cox} regression for doubly truncated data.
	{\em Biometrics} {\bfseries 74}, 481--487.
	\item [] Mart\' inez-Camblor, P. and Corral, N. (2012).
	A general bootstrap algorithm for hypothesis testing.
	{\em Journal of Statistical Planning and Inference} {\bfseries 142}, 589--600.
	\item [] Mart\' inez-Camblor, P. and de U\~na-\' Alvarez, J. (2009).
	Non-parametric k-sample tests: 
	{D}ensity functions vs distribution functions.
	{\em Computational Statistics and Data Analysis} {\bfseries 53}, 3344--3357.
	\item [] Moreira, C. and de U\~na-\' Alvarez, J. (2010).
	Bootstrapping the {NPMLE} for doubly truncated data.
	{\em Journal of Nonparametric Statistics} {\bfseries 22}, 657--583.
	\item [] Moreira, C., de U\~na-\' Alvarez, J. and Braekers, R. (2021).
	Nonparametric estimation of a
	distribution function from doubly truncated data under dependence.
	{\em Computational Statistics} {\bfseries 36}, 1693--1720.
	\item [] Moreira, C., de U\~na-\' Alvarez, J. and Crujeiras, R. (2022).
	\texttt{DTDA}: Doubly Truncated Data Analysis.
	{\em \texttt{R} package version 3.0.1}, https://CRAN.R-project.org/package=DTDA.
	\item [] Moreira, C., de U\~na-\' Alvarez, J. and Meira-Machado, L. (2016).
	Nonparametric regression with doubly truncated data.
	{\em Computational Statistics and Data Analysis} {\bfseries 93}, 294--307.
	\item [] Moreira, C., de U\~na-\' Alvarez, J. and Van Keilegom, I. (2014).
	Goodness-of-fit tests for a semiparametric model under random double truncation.
	{\em Computational Statistics} {\bfseries 29}, 1365--1379.
	\item [] Moreira, C. and Van Keilegom, I. (2013).
	Bandwidth selection for kernel density estimation with doubly truncated data.
	{\em Computational Statistics and Data Analysis} {\bfseries 61}, 107--123.
	\item [] Rennert, L. (2018).
	\texttt{SurvTrunc}: Analysis of Doubly Truncated Data.
	{\em \texttt{R} package version 0.1.0}, https://CRAN.R-project.org/package=SurvTrunc.
	\item [] Rennert L. and Xie, S.X. (2018).
	Cox regression model with doubly truncated data.
	{\em Biometrics} {\bfseries 74}, 725--733.
	\item [] Rennert L. and Xie, S.X. (2019).
	Bias induced by ignoring double truncation inherent in autopsy-confirmed survival studies of neurodegenerative diseases.
	{\em Statistics in Medicine} {\bfseries 38}, 3599--3613.
	\item [] Shen, P.S. (2010). 
	Nonparametric analysis of doubly truncated data.
	{\em Annals
		of the Institute of Statistical Mathematics} {\bfseries 62}, 835--853.
	\item [] Shen, P.S. (2013). 
	A class of rank-based tests for doubly-truncated data.
	{\em TEST} {\bfseries 22}, 83--102.
	\item [] Turnbull, B.W. (1976). 
	The empirical distribution function with arbitrarily grouped, censored and truncated data.
	{\em Journal of the Royal Statistical Society, Series B} {\bfseries 38}, 290--295.
	\item [] Woodroofe, M. (1985). Estimating a distribution function with truncated data.
	{\em Annals of Statistics} {\bf 13}, 163--177.
	\item [] Xiao, J. and Hudgens, M. G. (2019). On nonparametric maximum likelihood estimation with double truncation.
	{\em Biometrika} {\bf 106}, 989--996.
	\item [] Ying, Z., Yu, W., Zhao, Z. and Zheng, M. (2020). Regression analysis of doubly truncated data.
	{\em Journal of the American Statistical Association} {\bf 115}, 810--821.
	\item [] Zhu, H. and Wang, M.C. (2012).
	Analysing bivariate survival data with interval sampling and application to
	cancer epidemiology.
	{\em Biometrika} {\bfseries 99}, 345--361.
	\item [] Zhu, H. and Wang, M.C. (2014).
	Nonparametric inference on bivariate survival data with
	interval sampling: association estimation and testing.
	{\em Biometrika} {\bfseries 101}, 519--533.
	\item [] Zhu, H. and Wang, M.C. (2015).
	A semi-stationary copula model approach for bivariate
	survival data with interval sampling.
	{\em International Journal of Biostatistics} {\bfseries 11}, 151--173.
\end{itemize}


\end{document}